\documentstyle[budapest]{article}  
\frompage{000} \topage{000}                                              
\def\rightmark{Antikaons in isospin-asymmetric nuclear medium} 
\def\leftmark{C.L.\ Korpa and M.F.M.\ Lutz}
 
\title{Self-consistent antikaon dynamics in\\ 
isospin-asymmetric nuclear medium 
} 
\authors{
{C.L.~Korpa$^1$ and M.F.M.~Lutz$^{2}$ %
}\\[2.812mm]
{\normalsize
\hspace*{-8pt}$^1$ Department of Theoretical Physics, 
University of P\'ecs\\ 
7624 P\'ecs, Hungary\\[0.2ex] 
\hspace*{-8pt}$^2$ Gesellschaft f\"ur Schwerionenforschung\\ 
64291 Darmstadt, Germany
}}
 
\abstract{We investigate properties of antikaons and 
hyperon resonances in isospin-asymmetric nuclear medium,
using a self-consistent, covariant scheme based on 
vacuum antikaon-nucleon scattering amplitude.}

 
\begin{document}
 
\maketitle
\setcounter{page}{1}

\section{Introduction}\label{intro}
From the time of first suggestion that kaons may
condense in neutron stars \cite{Kaplan86}, study of their 
properties attracted much interest 
\cite{Politzer91,Brown92,Brown94,Brown96,Yabu93,Koch94,Waas96,Waas97,Ohnishi97,Lutz98,Ramos00}.
Understanding kaon behavior in nuclear medium is also necessary for
description of kaonic atoms \cite{Friedmann94,Lutz01} and subtreshold 
production of kaon in heavy-ion reactions \cite{Laue99}. The latter 
requires transport model calculations using codes which can incorporate
consistently particles with finite width, which are under development
\cite{Ivanov00,Leupold00,Cassing00,Schaffner00}.

Softening of the antikaon spectral function in the nuclear medium
was already anticipated from the $K$-matrix analyses of the
antikaon-nucleon scattering \cite{Martin81}, which predicted 
considerable attraction in the subtreshold scattering amplitudes.
For reliable calculation of the in-medium antikaon spectral 
function it is necessary to have an improved understanding of 
vacuum antikaon-nucleon scattering (especially for the subtreshold
region). 

Treatment of the antikaon-nucleon scattering is made rather involved 
by the open inelastic $\pi \Sigma$ and $\pi \Lambda$ channels, as well
as by the influence of the s-wave $\Lambda(1405)$, p-wave $\Sigma(1385)$
and d-wave $\Lambda(1520)$ resonances close to the threshold.
Furthermore, relatively large errors in the low-energy data make the subtreshold
extrapolation uncertain. It is then essential to use theoretical 
constraints (chiral symmetry, causality, covariance) while
constructing the scattering amplitudes. Two approaches (\cite{Kaiser9597}
and \cite{Oset98}) in this direction still gave markedly different 
subtreshold scattering amplitudes, triggering a recent work 
\cite{Lutz02a} which
systematically includes s-, p- and d-waves with total angular momentum
up to $J=3/2$. The amplitudes obtained in that work are approximately
crossing symmetric (the kaon-nucleon and antikaon-nucleon amplitudes
match closely in the subtreshold region) and are consistent with 
chiral symmetry and covariance.

For calculation of in-medium antikaon spectral function a self-consistent
scheme is necessary \cite{Lutz98}, since the effect of the modified
antikaon propagation on in-medium scattering process (for example on
the $\Lambda(1405)$) is very important. 
Recently a novel, covariant framework was developed \cite{Lutz02b}
in which the self consistency was implemented in terms of the vacuum
meson-nucleon scattering amplitudes. The method correctly takes
into account the in-medium mixing of s-, p- and d-waves,   
and was used to study
antikaon and hyperon-resonance properties in isospin-symmetric nuclear
medium \cite{Lutz02b}. 

In section 2.\ we extend the formalism 
to isospin-asymmetric medium with arbitrary neutron and proton desities.
In section 3.\ we present numerical results corresponding to neutron excess
typical for lead and also for neutron matter.

\section{Formalism}\label{form}  
We first briefly recall the self consistent and covariant many-body framework
introduced in \cite{Lutz02b}.
The vacuum on-shell antikaon-nucleon scattering amplitude is
\begin{eqnarray}
&&\langle {\bar K} ^{j}(\bar q)\,N(\bar p)|\,T\,| {\bar K} ^{i}(q)\,N(p) \rangle
=(2\pi)^4\,\delta^4(q+p-\bar q-\bar p)\,
\nonumber\\
&& \qquad \qquad   \qquad \qquad  \qquad \qquad \times \,\bar u(\bar p)\,
T^{ij}_{{\bar K}  N \rightarrow {\bar K}  N}(\bar q,\bar p ; q,p)\,u(p) \,,
\label{on-shell-scattering}
\end{eqnarray}
where the delta-function guarantees energy-momentum conservation and $u(p)$ is the
nucleon isospin-doublet spinor.
In quantum field theory the scattering amplitudes $T_{\bar{K}  N}$ 
follow as
solution of the Bethe-Salpeter matrix equation
\begin{eqnarray}
T(\bar k ,k ;w ) &=& K(\bar k ,k ;w )
+\int \frac{d^4l}{(2\pi)^4}\,K(\bar k , l;w )\, G(l;w)\,T(l,k;w )\;,
\nonumber\\
G(l;w)&=&-i\,S_N({\textstyle
{1\over 2}}\,w+l)\,D_{\bar{K} }({\textstyle {1\over 2}}\,w-l) \,,
\nonumber\\
w &=& p+q = \bar p+\bar q\,,
\quad k= \frac{1}{2}\,(p-q)\,,\quad
\bar k =\frac{1}{2}\,(\bar p-\bar q)\,,
\label{BS-eq}
\end{eqnarray}
in terms of the Bethe-Salpeter kernel $K(\bar k,k;w)$, the free space nucleon
propagator $S_N(p)=1/(p\!\!\!/-m_N+i\,\epsilon)$ and
kaon propagator $D_{\bar{K}}(q)=1/(q^2-m_K^2+i\,\epsilon)$.

The scattering process is readily generalized from the
vacuum to the nuclear matter case. In compact notation:
\begin{eqnarray}
&& {\mathcal T} = {\mathcal K} + {\mathcal K} \cdot {\mathcal G} \cdot {\mathcal T}  \;,\quad
{\mathcal T} = {\mathcal T}(\bar k,k; w,u) \;, \quad {\mathcal G} = {\mathcal G} (l;w,u) \,,
\label{hatt}
\end{eqnarray}
where the  in-medium scattering amplitude ${\mathcal T}(\bar k,k;w,u)$ and the two-particle
propagator ${\mathcal G}(l;w,u)$ depend now on the 4-velocity $u_\mu$
characterizing the nuclear matter frame. For nuclear matter moving with a velocity
$\vec v$:
\begin{equation}
u_\mu =\left(\frac{1}{\sqrt{1-\vec v\,^2/c^2}},\frac{\vec v/c}
{\sqrt{1-\vec v\,^2/c^2}}\right)
\;, \quad u^2 =1\,.
\end{equation}
The in-medium
two-particle propagator ${\mathcal G}$ is given by
\begin{eqnarray}
&& \Delta S_N (p,u) = 2\,\pi\,i\,\Theta \Big(p\cdot u \Big)\,
\delta(p^2-m_N^2)\,\big( p\!\!\! / +m_N \big)\,
\Theta \big(k_F^2+m_N^2-(u\cdot p)^2\big)\,,
\nonumber\\
&&{\mathcal S}_N(p,u) = S_N(p)+ \Delta S_N(p,u)\,, \quad
{\mathcal D}_{\bar K }(q,u)=\frac{1}{q^2-m_K^2-\Pi_{\bar K }(q,u)} \;,
\nonumber\\
&& {\mathcal G}(l;w,u) = -i\,{\mathcal S}_N({\textstyle
{1\over 2}}\,w+l,u)\,{\mathcal D}_{\bar K }({\textstyle {1\over 2}}\,w-l,u)  \;,
\label{hatg}
\end{eqnarray}
where the Fermi momentum $k_F$ is different for neutrons and protons.

The antikaon self energy $\Pi_{\bar K }(q,u)$ is
evaluated self consistently in terms of the relevant in-medium scattering
amplitudes
\begin{equation}
\Pi_{\bar K }(q,u) =  2\,{\rm{Tr}}\, \int \frac{d^4p}{(2\pi)^4}\,i\,\Delta
S_N(p,u)\,   \bar {\mathcal T}_{\bar K N}\big({\textstyle{1\over
2}}\,(p-q), {\textstyle{1\over 2}}\,(p-q);p+q,u \big).
\label{k-self}
\end{equation}

The in-medium scattering amplitude $\bar{\mathcal T}$ is defined 
with respect to the free-space interaction kernel $K$, since
we study only the effect of change in the medium propagation
of nucleons and antikaons:
\begin{equation}
\hat {\mathcal T}= K+K\cdot {\mathcal G}\cdot \hat {\mathcal T}
= T+T\cdot \Delta {\mathcal G} \cdot \hat {\mathcal T}\;,\quad
\Delta {\mathcal G}\equiv {\mathcal G}-G\;. \label{rewrite}
\end{equation}
 
The vacuum scattering amplitude can be decomposed systematically
into covariant projectors $Y_n^{(\pm)}(\bar q,q;w)$ with good
angular momentum and parity \cite{Lutz02a}. Including the 
leading terms with $J=1/2$ and $J=3/2$ 
(as was done in the numerical part of ref.\ 
\cite{Lutz02a}), means considering s-, p- and d-waves (the latter 
only with $J=3/2$). The relevant projectors are: for s-wave $Y_0^{(+)}$,
for p-wave with $J=1/2$ $Y_0^{(-)}$, for p-wave with $J=3/2$
$Y_1^{(+)}$ and for d-wave with $J=3/2$ $Y_1^{(-)}$, given by
\begin{eqnarray*}
Y_0^{(\pm )}(\bar q,q;w) &=& \frac{1}{2}\,\left( \frac{w\!\!\! /}
{\sqrt{w^2}}\pm 1 \right)\;,
\nonumber\\
Y_1^{(\pm)}(\bar q,q;w) &=& \frac{3}{2}\left( \frac{q\!\!\! /}{\sqrt{w^2}}\pm 1 \right)
\left(\frac{(\bar q \cdot w )(w \cdot q)}{w^2} -\big( \bar q\cdot q\big)\right)
\end{eqnarray*}
\vspace{-2mm}
\begin{equation}
-\frac{1}{2}\,\Bigg(  \bar q \!\!\! / -\frac{w\cdot \bar q}{w^2}\,w\!\!\! / 
\Bigg)
\Bigg(\frac{w\!\!\! /}{\sqrt{w^2}}\mp 1\Bigg)\,
\Bigg( q\!\!\! / -\frac{w\cdot q}{w^2}\,w\!\!\! / \Bigg)\;.
\label{vproj}
\end{equation}

In the presence of nuclear matter the scattering amplitude decomposition
is more complicated, due to appearance of another 4-vector, the medium
4-velocity $u_\mu$. The generalization of the projector algebra 
to this case has been worked out in ref.\ \cite{Lutz02b}. The physical 
reason for the presence of larger number of projectors is that the 
partial waves mix, thus necessitating introduction of the off-diagonal
matrix elements to the diagonal ones of eq.\ (\ref{vproj}). 

The mixing of partial waves happens in the following way. If the total
momentum of the nucleon and antikaon is zero in the rest frame of the
medium, there is no
mixing at all. In this case the rotational symmetry of the medium,
with respect to an observer who is at rest in it, assures that the
total angular momentum $J$ is a good quantum number which, together with
parity, prevents mixing of partial waves. 

If the observer moves with
respect to the medium, i.e.\ $\vec w\neq 0$ in the rest frame of
the medium, not all components of the total angular 
momentum are good quantum numbers, since only rotations about an
axis defined by the momentum of the observer are still a symmetry. This
means that only helicity remains a good quantum number and that states 
with different helicity do not mix (actually, what matters in our case 
is only the absolute value of the helicity). This picture is borne out
by the projector algebra worked out in \cite{Lutz02b}, where it was 
shown that the algebra
decomposes into two subalgebras. One corresponds to elements of a 
2$\times$2 matrix, where the diagonal elements are the amplitudes of the
two partial waves with $J=3/2$. This obviously describes the mixing of the 
helicity-3/2 states. The other subalgebra, after elimination of zero
entries, corresponds to a 4$\times$4 matrix, whose diagonal elements 
are the amplitudes of all four partial waves. This describes the
mixing of helicity-1/2 states, present in all four considered partial
waves.

We now turn to considering the effects of isospin asymmetry of the
nuclear medium. Consequently, we have to evaluate separately the
self energy of the neutral antikaon and the negative kaon, using
appropriate in-medium scattering amplitudes in expression 
(\ref{k-self}). In each case there are two integrations, over the
neutron and proton Fermi sea. 

We chose to work in the particle bases, as opposed to the isospin one.
In the former the loop integrals, containing as integrand the 
propagators of a nucleon and an antikaon, are obvously diagonal.
There is, of course, mixing in some vacuum scattering amplitudes, 
which are given as:
\[
T_{p\bar K^0\rightarrow p\bar K^0}=
T_{n K^-\rightarrow n K^-}=T^{(1)},\;\;
T_{n\bar K^0\rightarrow n\bar K^0}=
T_{p K^-\rightarrow p K^-}=\frac{1}{2}(T^{(0)}+T^{(1)}),
\]
\vspace{-3mm}
\begin{equation}
T_{p K^-\rightarrow n\bar K^0}=
T_{n \bar K^0\rightarrow p K^-}=\frac{1}{2}(T^{(0)}-T^{(1)}),
\end{equation}
where $T^{(0)}$ ($T^{(1)}$) is the isospin-0 (isospin-1) amplitude.

\section{Numerical results}\label{num}
The computational scheme outlined in the previous section
requires as input only the vacuum antikaon-nucleon scattering 
amplitudes for s-, p- and d-waves. The amplitudes recently
obtained in ref.\ \cite{Lutz02a} are especially well suited for
this purpose, since they systematically incorporate constraints
from chiral symmetry, crossing symmetry and causality, and 
describe the antikaon-nucleon scattering quantitatively up to 
kaon laboratory momentum of 500 MeV (qualitatively up to
1000 MeV).

We use an iterative procedure to solve the eq.\ (\ref{rewrite}),
starting from the calculation of antikaon self energies from
vacuum amplitudes. Then we calculate the loop integrals, i.e.\
$\Delta{\mathcal G}$, which is used to compute the in-medium scattering
amplitudes. The elf energies are then computed from the latter ones
and the procedure is repeated until convergence is achieved (4--5
iterations are enough).

\subsection{Antikaon spectral function in nuclear medium}\label{kaon}
In fig.\ 1 we show the antikaon spectral functions in neutron 
matter of density $\rho=0.17\;{\rm fm}^{-3}$, for different kaon 
momenta. The effect of the medium is more pronounced on the 
$\bar K^0$, since protons are not present and thus the $K^-$ 
can not produce the $\Lambda(1405)$. 

\vspace{0mm}
\epsfxsize=14cm
\centerline{\epsffile{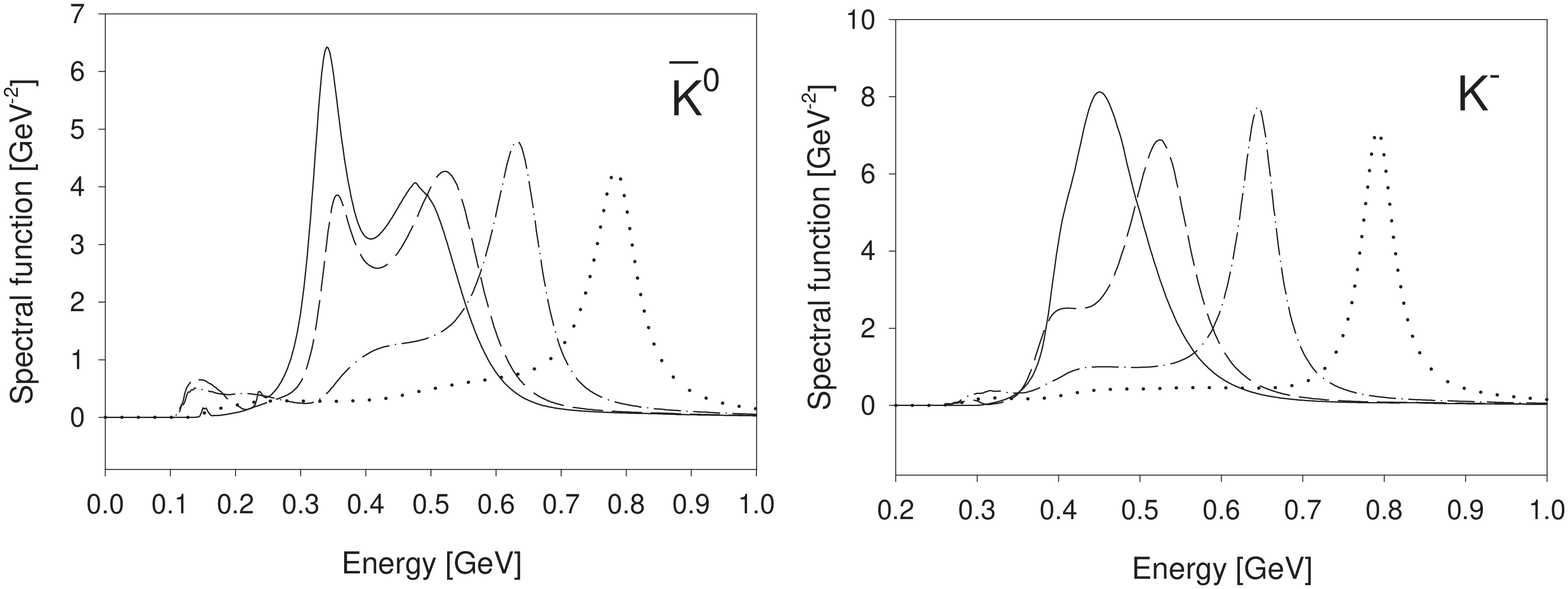}}

\vspace{-3mm}
\begin{center}
Fig.\ 1. Antikaon spectral functions in neutron matter of
density $\rho=0.17\;{\rm fm}^{-3}$, for different momenta:
0 (full line), 200 MeV (dashed line), 400 MeV (dash-dot line),
600 MeV (dotted line).
\end{center}

In fig.\ 2 the antikaon spectral functions are shown for the
case of two lead nuclei superimposed, i.e. for neutron and 
proton densities double that of the central part of lead 
nucleus (we used the following Fermi momenta: for protons
$p_{F_p}=310.8\;{\rm MeV}$ and for neutrons 
$p_{F_n}=355.8\;{\rm MeV}$). We observe a dramatic broadening 
at small to intermediate momenta (up to about 500 MeV),which
means that the quasiparticle approximation is inadequate. Even
at larger momenta, 600--700 MeV, the broadening is quite
significant and the spectral function is nonzero (although
small) in the low-energy region.
Again, 
the neutral antikaon is more affected, but the difference
is not pronounced. Both spectral functions show nonzero support at 
small energy (starting from around 90 MeV) and that may have
non-negligible effects in heavy-ion collisions.

\vspace{0mm}
\epsfxsize=14cm
\centerline{\epsffile{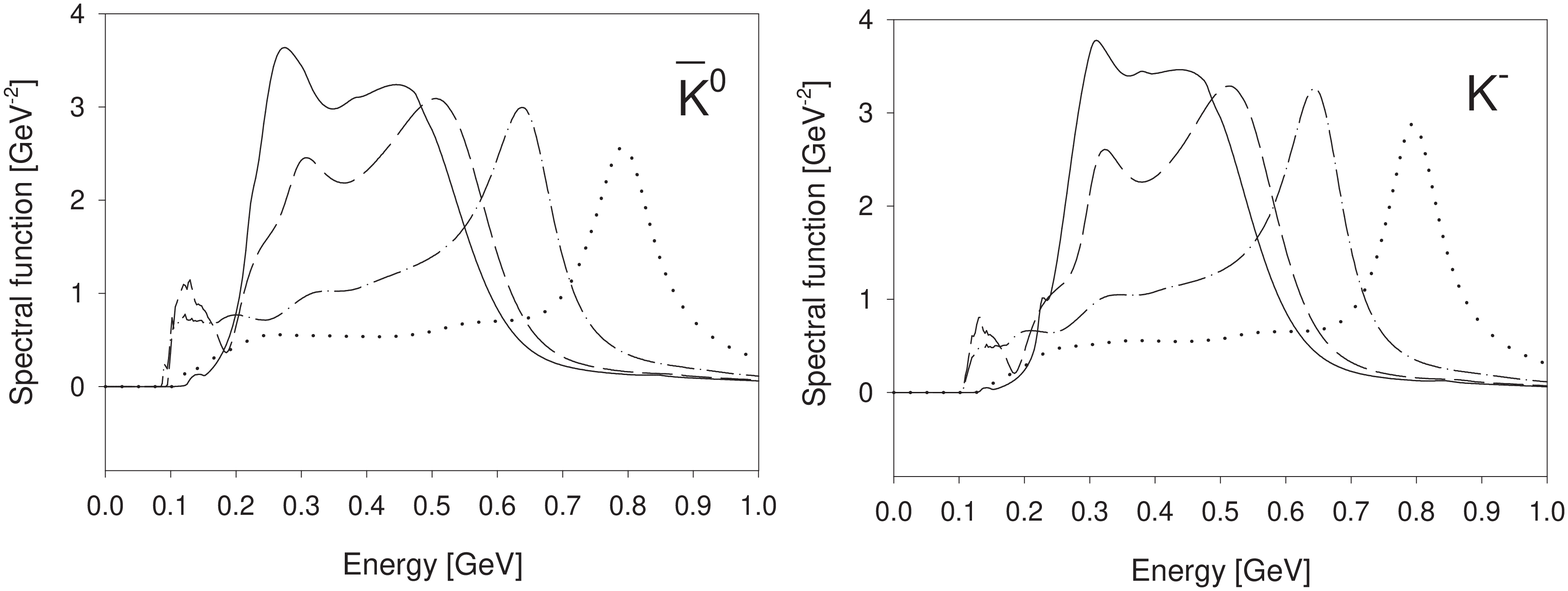}}

\vspace{-3mm}
\begin{center}
Fig.\ 2. Antikaon spectral functions in the medium with proton
and neutron density double that of the central part of lead nucleus,
for different momenta:
0 (full line), 200 MeV (dashed line), 400 MeV (dash-dot line),
600 MeV (dotted line).
\end{center}

\subsection{Hyperon resonances in nuclear medium}\label{hyperon}
From the in-medim scattering amplitudes we can infer the changes
in the hyperon resonances playing a role in the considered s-, p- 
and d-waves. The $\Lambda(1405)$ plays a dominant role in the 
$I=0$ s-wave and is modified considerably in isospin-symmetric
medium \cite{Lutz02b}. In fig.\ 3 we show the real and imaginary 
part of the corresponding scattering amplitude in neutron matter 
and, for comparison, vacuum.

\vspace{0mm}
\epsfxsize=8cm
\centerline{\epsffile{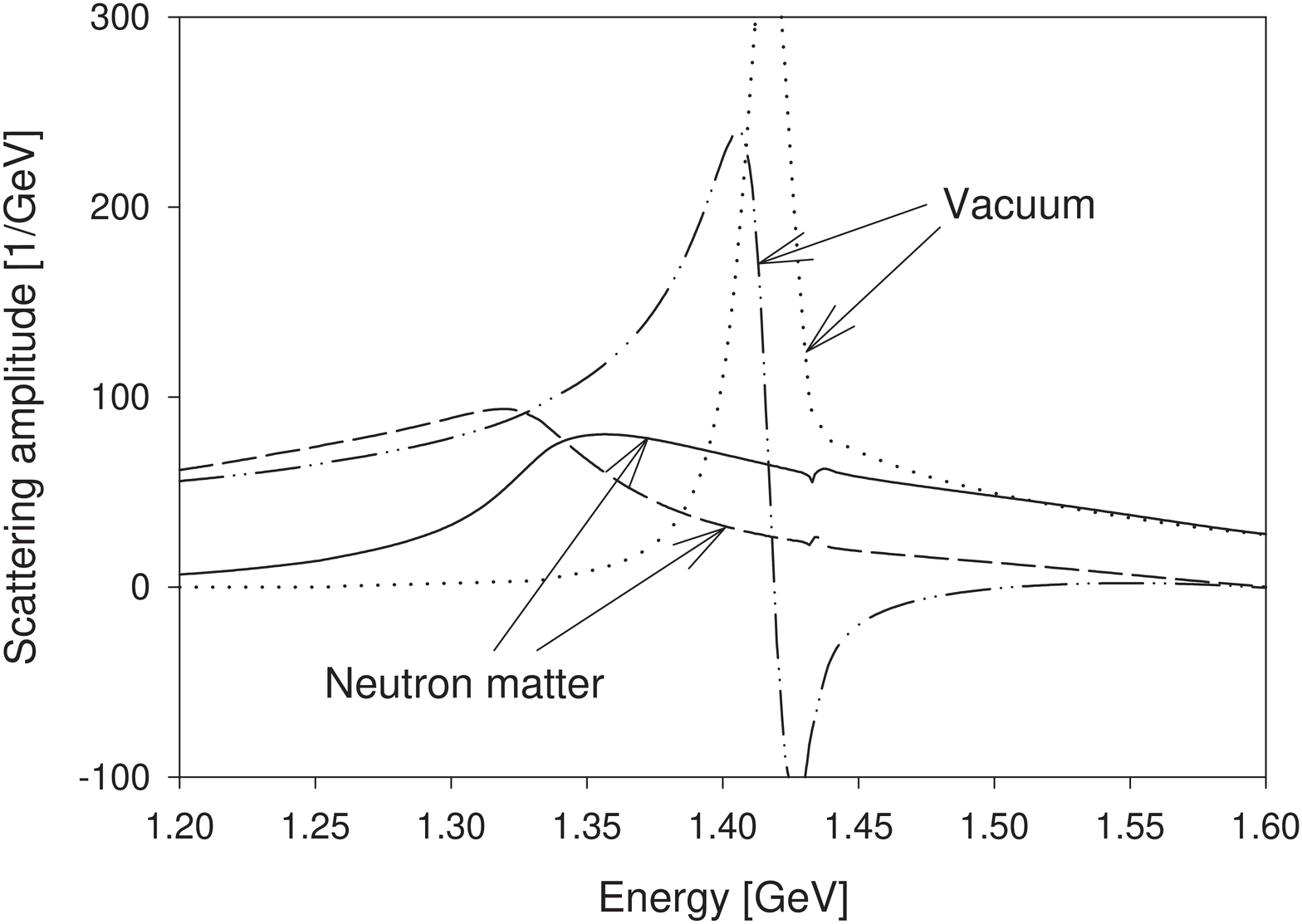}}

\vspace{-3mm}
\begin{center}
Fig.\ 3. The $\Lambda(1405)$ at rest in neutron matter of
density $\rho=0.17\;{\rm fm}^{-3}$. The full (dashed) line
shows the imaginary (real) part of the relevant scattering
amplitude, while the dotted (dash-dot-dot) line shows the
imaginary (real) part of the amplitude in vacuum.
\end{center}

For lead (where the neutron number is 50\% higher than the
proton number) the effect on $\Lambda(1405)$ is quite similar 
to the one in isospin-symmetric medium with the same nucleon 
density, as shown in ref.\ \cite{Lutz02b}. However, increasing
the proton and neutron density by a factor of two produces even
more pronounced broadening than the one shown in fig.\ 3.

It may be of interest to look at the splitting of the $\Sigma(1385)$ 
in isospin-asymmetric medium. Figs.\ 4 and 5 show the 
imaginry part of the corresponding amplitudes for zero total 
momentum (fig.\ 4) and 400 MeV total momentum (fig.\ 5).

\vspace{4mm}
\epsfxsize=13.cm
\centerline{\epsffile{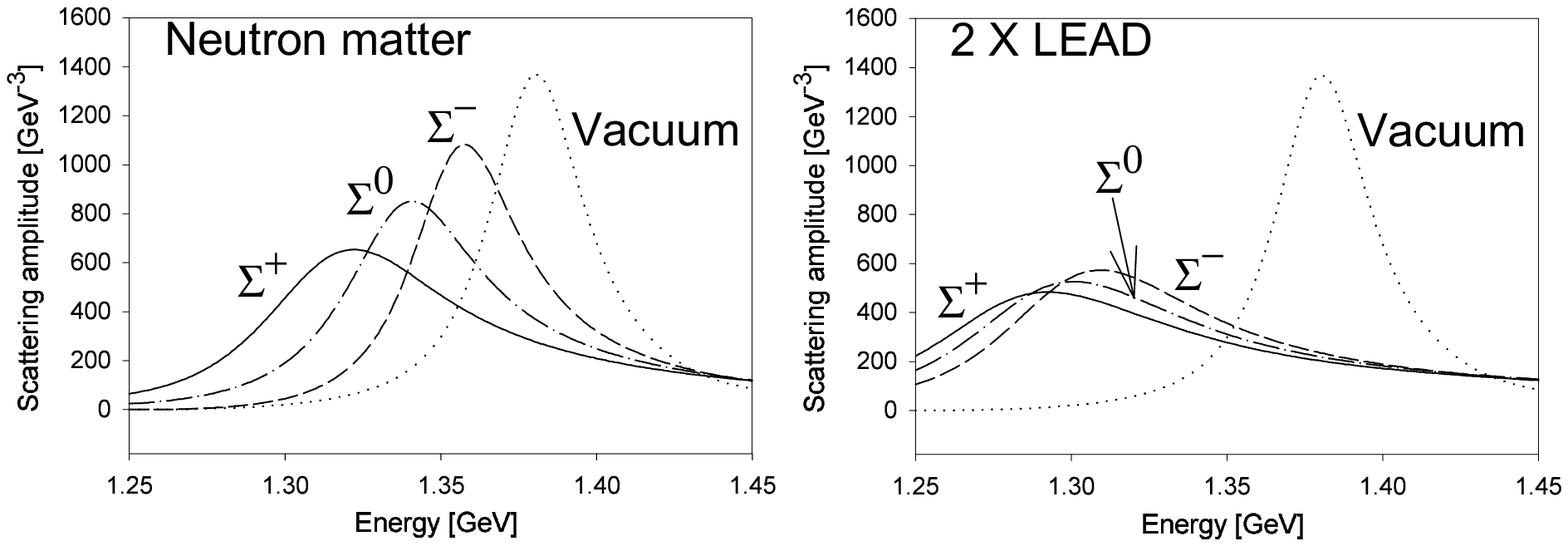}}

\vspace{-3mm}
\begin{center}
Fig.\ 4. The $\Sigma(1385)$ at rest in neutron matter of
density $\rho=0.17\;{\rm fm}^{-3}$ (left figure) 
and nuclear matter double the lead density (right figure). 
The full line
shows the imaginary part of the scattering amplitude
corresponding to the $\Sigma^+$, dashed line to $\Sigma^-$ and
dash-dot line to $\Sigma^0$. The vacuum amplitude is shown
for comparison by dotted line.
\end{center}

\vspace{4mm}
\epsfxsize=13.cm
\centerline{\epsffile{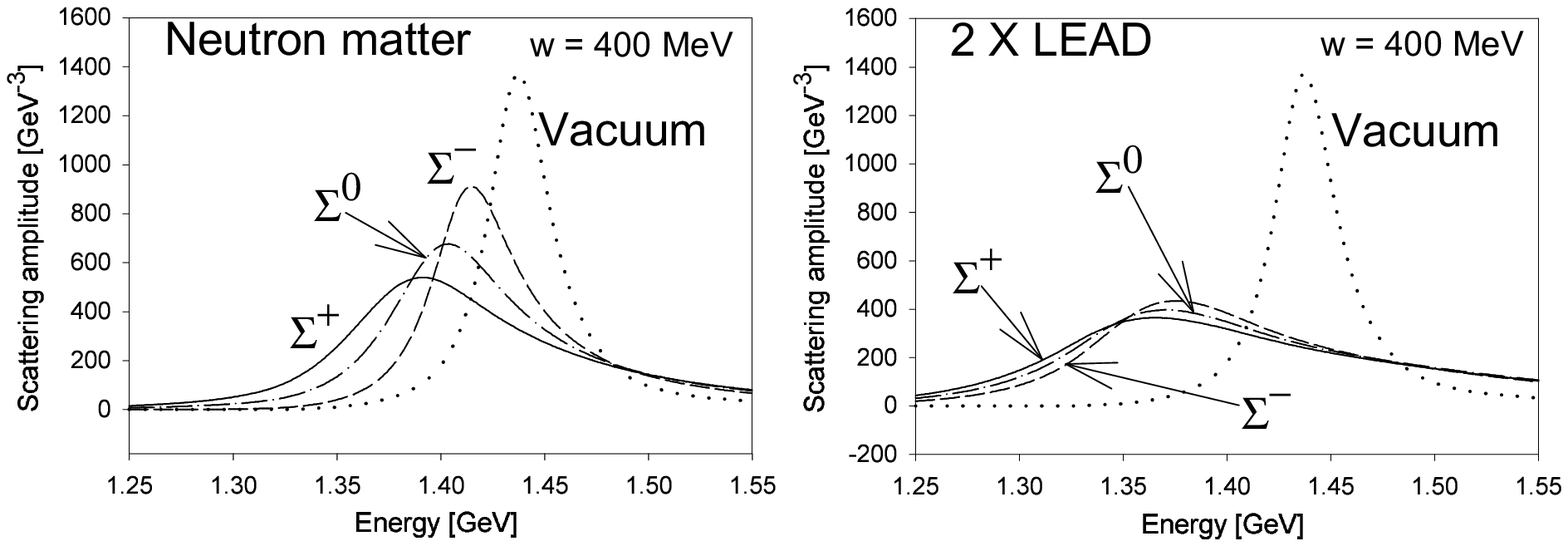}}

\vspace{-1mm}
\begin{center}
Fig.\ 5. The same as fig.\ 4, with only difference that the 
$\Sigma(1385)$ moves with momentum $|\vec w|=400\;{\rm{MeV}}$ in
the medium.
\end{center}

The maxima of the resonances usually obey with good accuracy
the vacuum relation $\sqrt{m_*^2+\vec w\,^2}$, where $m_*$ is the
position for the resonance at rest. However, as seen in fig.\ 5,
the $\Sigma(1385)$ resonances are closer to each other at
$|\vec w|=400\;{\rm MeV}$, than the above relation would suggest.

The $\Sigma(1195)$ in neutron matter shows a splitting which is smaller than that
of the $\Sigma(1385)$. The position of the maximum for the 
$\Sigma^+$ at rest is 1180 MeV, while the $\Sigma^-$ is at 
1192 MeV, with the $\Sigma^0$ being just in the middle between
the charged states.

Strong medium modification characterizes also the other 
resonances playing part in the antikaon-nucleon scattering, 
for the case of presented densities. Exception is the $\Sigma(1690)$
which suffers only a minor broadening, probably as a consequence 
of small branching fraction into antikaon-nucleon channel.

In conclusion, we extended the recently developed method
for self-consistent analysis of antikaons and hyperon resonances
in isospin-symmetric nuclear medium to the isospin-asymmetric one.
In neutron matter we observe a stronger medium modification of the
$\bar K^0$ as compared to the $K^-$. The magnitude of the 
medium effects on the hyperon resonances is similar to the 
isospin-symmetric case, but with pronounced splitting of the
$\Sigma(1385)$ and a smaller one for the $\Sigma(1195)$.

\section*{Acknowledgement}
This research was supported in part by the
Hungarian Research Foundation (OTKA) grant T030855.


\vfill\eject
\end{document}